\documentclass[bibyear]{aa} 

\usepackage{graphicx}
\usepackage{amsmath}
\usepackage{float} 
\usepackage{caption}

\usepackage{txfonts}

\begin{document}

  \title{Theoretical tidal evolution constants for stellar  models 
 from the pre-main  sequence  to the white dwarf stage}

   \subtitle{Apsidal motion constants,    
        moment of inertia, and gravitational potential energy  }
   
\author{A. Claret\inst{1, 2}}
   \offprints{A. Claret, e-mail:claret@iaa.es. Tables 1–69 are 
        only available at the CDS via anonymous ftp
        to cdsarc.u-strasbg.fr (130.79.128.5) or via http://cdsarc.
        u-strasbg.fr/viz-bin/qcat?J/A+A/vol/page
     }
\institute{Instituto de Astrof\'{\i}sica de Andaluc\'{\i}a, 
        CSIC, Apartado 3004,  18080 Granada, Spain
        \and
     Dept. Física Teórica y del Cosmos, Universidad de Granada, 
     Campus de Fuentenueva s/n, 10871 Granada, Spain
}
            \date{Received; accepted; }
\abstract
       {} 
         {One of the most reliable means of studying the stellar interior is 
        through the apsidal motion in double line eclipsing binary systems since 
        these systems present errors in masses, radii, and effective temperatures of only a 
        few per cent. On the other hand, the theoretical values of the apsidal motion to 
        be compared with the observed values depend on the stellar masses 
        of the components and more strongly on their radii (fifth power). 
        The main objective of this work is to make available  grids of evolutionary stellar models 
        that, in addition to the traditional parameters (e.g. age, mass, log g, T$_{\rm eff}$), also 
        contain the necessary parameters for the theoretical study of apsidal motion 
        and tidal evolution. This  information is useful for the study of the 
        apsidal motion in eclipsing binaries and their tidal evolution, and can also be used 
        for the same purpose in exoplanetary systems.} 
        {All models were computed using the MESA package. We consider core overshooting for 
        models with masses $\ge$  1.2 M$_\odot$. For the amount of core overshooting we 
        adopted a recent  relationship for mass $\times$ core overshooting. We  adopted for the mixing-length parameter 
        $\alpha_{\rm MLT}$ the value  1.84 (the solar-calibrated value).  
        Mass loss was   taken into account in two evolutionary phases. The models were followed from the pre-main sequence phase to the white 
        dwarf  (WD) stage. }
   {The evolutionary models containing age, luminosity, log g, and Teff,   as well as 
        the first three   harmonics of the internal stellar structure (k$_2$, k$_3$, and  k$_4$), 
        the radius of gyration $\beta$ y, and the dimensionless variable $\alpha$, related to 
        gravitational potential energy, are presented in 69 tables covering three chemical 
        compositions: [Fe/H] = -0.50, 0.00, and 0.50.  Additional models with different input physics are available. }
   {}

   \keywords{stars: binaries: eclipsing; stars: evolution; stars: white dwarfs;
    stars: interiors;  stars: planetary systems}
   \titlerunning {Tidal evolution }
   \maketitle
%

\section{Introduction} 

Double line eclipsing binary systems (DLEBS) are the best sources for obtaining 
absolute stellar parameters with great precision. In addition, because of  their proximity, 
some effects may appear due to the interaction of the two components, such as mutual 
irradiation, tidal distortion, and mass exchange. DLEBS are very important in 
astrophysics because the perturbations due to the proximity of the two
components act as probes and make it possible to investigate in detail 
the evolution and in some particular cases their internal structure. In this sense, such
perturbations play a very similar role to that of the usual techniques of physics labs 
  in which objects are perturbed applying for example a magnetic and/or electric field 
and studying its behaviour under the action of the applied perturbations. 

In the case of DLEBS the presence of the companion changes the
gravitational field of both, which   affects the equilibrium configuration of both components (effect
of tides). This alteration is responsible for the loss of 
 spherical symmetry of the binary components and it depends on the internal structure of the 
components. The two stars can also be distorted by the effect of the 
rotation that tends to flatten them on the poles.  From the theoretical point of view, it 
is possible  to describe such  distortions as a function of the internal 
structure of both stars. The orbit of this pair of stars will not be Keplerian 
because the orbital elements will be  functions of time, in particular  the argument of 
  periastron $\omega$. In general, there are three physical phenomena that can give rise to 
apsidal motion:  the loss of the spherical symmetry by distortions;  the presence of a third body; 
and a relativistic effect,  the best known 
example being  the advance of  perihelion of the planet Mercury. 

On the other hand,  earlier comparisons  between the internal structure constants 
 derived from the observed apsidal motions were reported a long time ago; they  
 indicate  that stellar structure is more centrally concentrated in 
mass than those extracted from the stellar evolutionary models  (see e.g.  
  Introduction in Claret\&Giménez 1993). Such discrepancies 
 were partially resolved later by Claret\&Gim\'enez (1993, 2010) considering 
new times of minima,  new opacity tables, and  core overshooting. 

For several years the apsidal motion of DI Her was a serious problem 
since the comparison between the theoretical calculations and the observed value of  
$\dot\omega$ differed by almost 500\%. Various mechanisms were invoked to explain 
such a discrepancy, including alternative 
theories of gravitation. For the confrontation of  theory and observational data,  Claret (1998) 
analysed  some  aspects of the apsidal motion of DI Her. The  main conclusion of that paper  
was that   an alternative theory of gravitation was not necessary to explain the observational 
value of the apsidal motion.  Finally the case of DI Her  was solved observationally through 
the Rossiter-McLaughlin effect  by Albrecht et al. (2009). Later Claret et al. (2010) using 
the data obtained by Albrecht et al. (2009), mainly those related to the Rossiter-McLaughlin effect, 
found a good agreement between the theoretical value of k$_2$ and its observational counterpart. 
More recently, Lang, Winn, and Albrecht (2022), using TESS data combined with previous observations,  
obtained a significant result since  the three-dimensional spin directions of the two 
components of DI Her could be determined. With these data these authors have found  
good agreement between   k$_{\rm 2theo}$, provided by Claret et al. (2021),  and its 
observational counterpart.

To the best of our knowledge, the last systematic comparison between   theoretical and 
observed values of apsidal motion rates was carried out by Claret et al. (2021) who 
used an observational sample of 27 selected DLEBS to compare the theoretical values 
of k$_2$ with their observational counterparts.  
These authors have used minimum times  extracted from the light curves provided 
by the Transiting Exoplanet Survey 
Satellite (TESS). Very good agreement has been found between the theoretically 
predicted values and their observational counterparts, including the troublesome 
case of DI Her.

Another very important contribution to the study of apsidal motion came from a group from 
the Astronomical Institute at Charles University.  Zasche\&Wolf (2019) investigated 21 
eccentric eclipsing binaries (early-type) located in the Small Magellanic Cloud and determined 
their apsidal motions and analysed their respective light curves. 
More recently Zasche et al. (2021) present an extensive sample of 162 early-type binary 
systems showing apsidal motion located in the Large Magellanic Cloud. This point is 
particularly important given that light curves and apsidal motion modelling were carried out 
for the first time for several systems simultaneously and in an environment with a  chemical 
composition different from  solar (for a more recent reference on apsidal motion measurements,   
see Zasche et al. 2023). 

The comments in the previous paragraphs refer mainly to DLEBS that are still on the 
main sequence  or close to it.   Burdge 
et al. (2019) studied the orbital decay of compact stars in a hydrogen-poor low-mass white dwarf (WD),  
FJ053332.05+020911.6. One of the components of this system exhibits 
ellipsoidal variations due to tidal distortions.  The estimated mass for 
this component (PTF J0533+0209B) is of the order  of 0.20 M$_\odot$. 
We note that the gravity-darkening  effect must also be taken into account 
for compact stars distorted by tides and/or rotation (Claret 2021).
Until then we  computed our evolutionary stellar models containing internal structure 
constants only up to the giant phases.  For the case of PTFJ053332.05+020911.6 
and other similar systems, where at least one  WD has been detected, we   
decided to extend our grids  from the pre-main sequence (PMS)  to the  WD phase. 
This is the main objective of the present paper.

\section {Stellar evolutionary models: Apsidal motion internal constants, 
momentum of inertia, and  gravitational potential energy}

 The evolutionary  tracks were computed using  the Modules for Experiments in Stellar 
Astrophysics package (MESA; see  Paxton et al. 2011, 2013, 2015;  v7385). 
We introduced a subroutine to compute the apsidal motion constants (k$_2$, k$_3$, k$_4$), 
the moment of inertia, and the gravitational potential energy. In this paper  we do 
not consider directly the effects of  rotation. The adopted mixing-length parameter 
$\alpha_{\rm MLT}$ was  1.84 (the solar-calibrated value; Torres et al. 2015). 
However, the $\alpha_{\rm MLT}$  parameter seems to depend on  the evolutionary status 
and/or metallicity,  as shown by  Magic et al. (2015) using   3D simulations. 
As commented in Claret (2019), it is not easy to  compare these results with  those coming 
from MESA, due to the different input physics,  for example  the equation of state 
and opacities. For the  opacities we  adopted the  element mixture given by 
Asplund et al. (2009). The helium content follows  the  enrichment law 
Y = Y$_p$ + 1.67 Z, where Y$_p$ is the primordial helium content (Ade et al. 2016). 

The mass range covers the interval from 0.2   to  8.0 M$_{\odot}$ for three chemical 
compositions: [Fe/H] -0.5, 0.00, and +0.50. The grids for the  two extra chemical compositions,  
[Fe/H ]=-0.50 and 0.50, were computed to take into account observational errors in [Fe/H]  
for systems located in the solar environment. 

As commented in the Introduction, the evolutionary tracks were computed from the PMS  up  to the 
WD stage. We adopted the following scheme for mass loss: for the  
interval 0.2--1.8 M$_{\odot}$ we  followed the recipe by Reimers (1977)  with $\eta_R$ = 0.1 
and for the  AGB scheme  we adopted the formalism by  Blocker (1995) with $\eta_B$ = 10.0.  
For models more massive than 1.8 M$_{\odot}$  we assumed  $\eta_R$ = 0.1 and  $\eta_B$ = 30.0.  
The adopted wind switch RGB-AGB  was 1.0$\times10^{-4}$.  

Convective core overshooting  was considered for models with stellar mass higher than 
or equal to 1.2 M$_{\odot}$. In this paper we adopt the  diffusive approximation, 
represented  by the free parameter f$_{ov}$ (Freytag et al. 1996  and 
Herwig et al. 1997).  The diffusion  coefficient in the overshooting region is 
given by the expression ${D_{ov}} = D_o exp\left({{-2z\over{H_{\nu}}}}\right)$, where   
D$_o$ is the diffusion coefficient at the convective boundary, $z$ is the 
geometric distance from the edge of the convective zone, H$_{\nu}$ is the velocity 
scale-height at the  convective boundary expressed as  H$_{\nu}$  = f$_{ov}$ H$_p$, 
and the coefficient  f$_{ov}$ is a free parameter that governs the width of  
the overshooting layer.  It  is known that models computed adopting  core overshooting 
are more centrally concentrated in mass than their standard  counterparts following 
Claret\&Giménez (1991). For the amount of core overshooting we adopted the 
relationship between the stellar mass and f$_{ov}$ derived by Claret\&Torres 
(2019), instead of adopting a single value of core overshooting  for the entire 
range of masses, as was done in the past.

\begin{figure}
        \includegraphics[height=9cm,width=6cm,angle=-90]{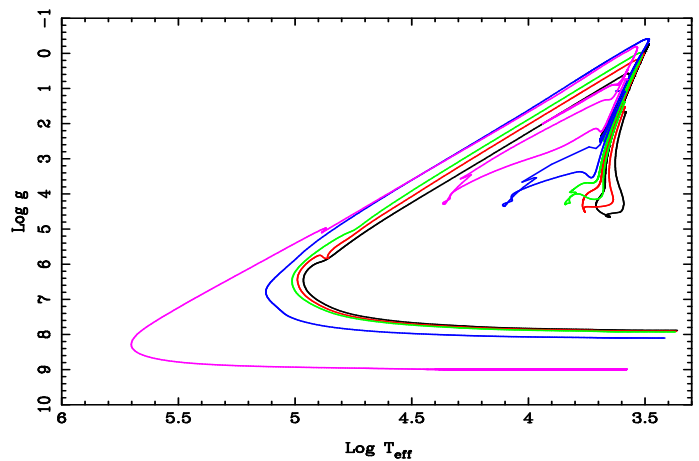}
        \caption{Hertzprung--Russell diagram for some models from the PMS to cooling WD stage. The 
        masses of the models  are (from right to 
        left)  0.7 (black), 1.0 (red), 1.4 (green), 3.0 (blue), and 8.0 (magenta) in solar units.  
        $\alpha_{\rm MLT} = 1.84$, [Fe/H] = 0.00. }
\end{figure}

\subsection {Apsidal motion constants: k$_2$, k$_3$, and k$_4$}

The theoretical  apsidal motion constants   k$_2$, k$_3$, and k$_4$   were  derived simultaneously  by integrating  the Radau equation using a 
fifth-order Runge-Kutta method, with a tolerance level of  10$^{-7}$:

\begin{eqnarray}
{a{\rm d}\eta_{j}\over {\rm d}a}+ {6\rho(a)\over\overline\rho(a)}{(\eta_{j}\!+\!1)}+
{\eta_{j}(\eta_{j}\!-\!1)} = {j(j+1}), \, j=2,3,4.
\end{eqnarray}

\noindent
Here the auxiliary parameter $\eta_j$  is given by 

\begin{eqnarray}
\eta_j \equiv {{a}\over{\epsilon_{j}}} {{\rm d}\epsilon_{j}\over{{\rm d}a}}.  
\end{eqnarray} 

\noindent
In Eq. 1,  $a$ denotes the mean radius of the stellar configuration, $\epsilon_j$  is  a  
measure of the deviation from sphericity, $\rho(a)$ is the 
mass density at the distance {\it a} from the centre of the configuration, and 
 $\overline\rho(a)$ is the mean mass density within a sphere of radius {\it a}.

The  apsidal motion constant of order $j$ is   given by 

\begin{eqnarray}
k_{j} = {{j +1 - \eta_{j}(R)}\over{2\left(j+\eta_{j}(R)\right)}}
,\end{eqnarray}

\noindent
where $\eta_{j}(R)$ indicates the values  of $\eta_{j}$ at the surface of the star. 
We note that these equations were  derived in the framework of static tides.  
For  the case of dynamic tides, we need to treat with more elaborated 
equations because the rate of static tides  is derived assuming that the orbital 
period is larger than the periods of the free oscillation modes.  
However, dynamic tides can significantly change  this scenario  due to the 
effects of the compressibility of the stellar fluid. This  is important in  
systems that are nearly  synchronized  
synchronism. In this case, for higher rotational angular velocities, additional
deviations due to resonances   appear if the forcing frequencies of the dynamic 
tides come into the range of the free oscillation modes of the component stars. 
The role  of   dynamical tides  was evaluated for some DLEBS by   
Claret\&Willems (2002), Willems\&Claret (2003), Claret\&Giménez  (2010), and 
more recently in  Claret et al. (2021).

As mentioned in the Introduction, our stellar evolutionary tracks  were computed without  
taking rotation into account. In order to evaluate the effects of rotation on the apsidal 
motion  constants,  a correction on the internal structure constants  was proposed by 
Claret (1999). This  correction  is  given by the  equation 

\begin{eqnarray}
\Delta   {\rm log k_2} \equiv {\rm log k_{ 2, standard}} - \lambda.
\end{eqnarray}
 
\noindent
Here {$\lambda=2V^{2}/(3gR)$,   where  $g$ is the surface gravity and $V$ is the  equatorial   
rotational velocity.

\begin{figure}
        \includegraphics[height=9cm,width=6cm,angle=-90]{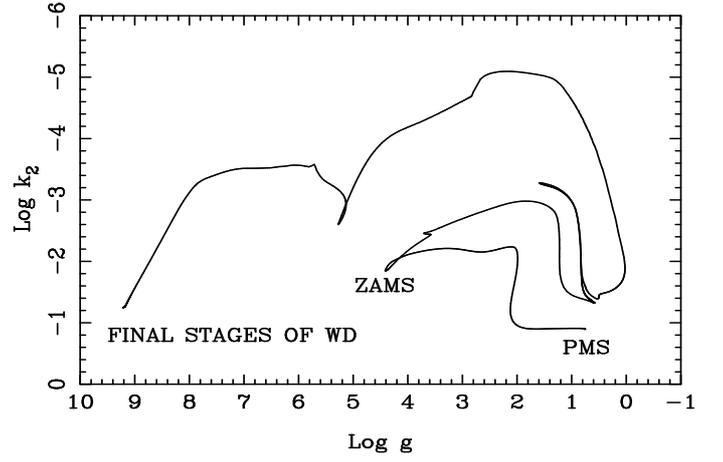}
        \caption{Behaviour of logk$_2$ as a function of log g. [Fe/H]=0.00, initial
         mass of 8.00  M$_{\odot}$.}
\end{figure}

\begin{figure}
        \includegraphics[height=9cm,width=6cm,angle=-90]{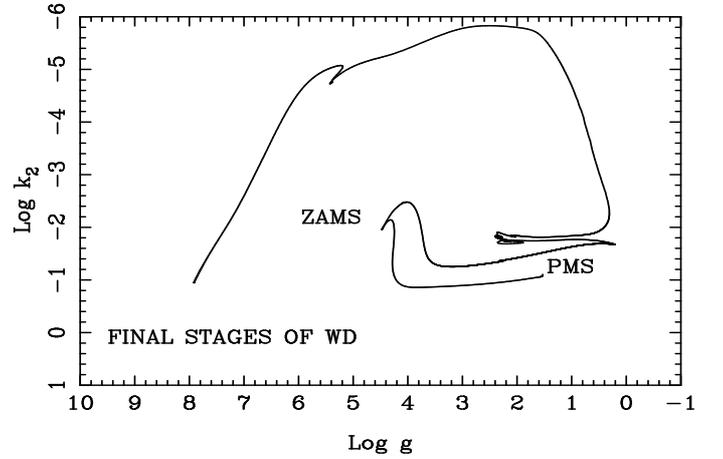}
        \caption{Same as  Figure 2,  but for an  initial
                mass of 1.00  M$_{\odot}$.}
\end{figure}

Figure 1 shows the Hertzprung--Russell diagram (HR) for some selected models:  0.70, 
1.00, 1.40, 3.00, and 8.00 M$_{\odot}$.  
In Figs. 2 and 3 we show the evolution of log k$_2$ as a function of log g 
for models with initial masses of 8.00 and 1.00  M$_{\odot}$, respectively. 
The behaviour of the two models is similar   when they reach the   
WD stage  (log k$_2$ is of the order of -1.00). This implies that, using simple 
models based on polytropes, the equivalent $n$ would be of the order of 2.1, where 
$n$ is the polytropic index. This data confirm the earlier calculations for WD 
using polytropes as input physics. We recall that for the case of non-relativistic electrons 
$n$ $\approx$ 1.5 and for the case of relativistic electrons $n$ = 3.0 this  index 
would be $\approx$ 2.0 (see Kopal (1959), pag. 35). 

The parameter k$_2$ is   applied in the studies of apsidal motion of DLEBS 
and/or exoplanets, and is also useful for computing tidal evolution.  For example, the 
differential equations that govern the tidal evolution depend not only on this parameter, 
but also on the radius of gyration (see Hut 1980, 1982). We can write the 
corresponding differential equations as

\begin{eqnarray}
        {de\over {dt}} = -{27 k_{2 1}\over{t_{F1}}} q (q +1) \left(R_1\over{A}\right)^8 
        {e \over{(1-e^2)^{13/2}}} \nonumber \\
        \left( f_3 - 11/18(1-e^2)^{3/2} 
        f_4 {\Omega_1\over{\omega}}\right),
\end{eqnarray}

\begin{eqnarray}
        {dA\over {dt}} = -{6 k_{2 1}\over{t_{F1}}} q (q +1) \left(R_1\over{A}\right)^8 
        {A \over{(1-e^2)^{15/2}}}  \nonumber \\ \left( f_1 - (1-e^2)^{3/2} 
        f_2 {\Omega_1\over{\omega}}\right),
\end{eqnarray}

\begin{eqnarray}
        {d\Omega_1\over {dt}} = {3 k_{2 1}\over{t_{F1} \beta_1^2}} q^2 
        \left(R_1\over{A}\right)^6 {\omega \over{(1-e^2)^{6}}} 
        \left( f_2 - (1-e^2)^{3/2} f_5 {\Omega_1\over{\omega}}\right),
\end{eqnarray}

\begin{eqnarray}
        {d\Omega_2\over {dt}} = {3 k_{2 2}\over{t_{F2} \beta_2^2}} q_2^2 
        \left(R_2\over{A}\right)^6 {\omega \over{(1-e^2)^{6}}} 
        \left( f_2 - (1-e^2)^{3/2} f_5 {\Omega_2\over{\omega}}\right).
\end{eqnarray}

\noindent 
In the above equations   $e$ represents the orbital eccentricity, A is the semi-major axis, 
M$_i$ is the mass of component $i$,  $\Omega_{i}$  is the angular velocity of the component $i$,  
$\omega$ is the mean orbital angular  velocity,  R$_i$ is the radius of the component 
$i$, q = M$_2$/M$_1$, q$_2$ = M$_1$/M$_2$, and    t$_{\rm Fi}$ is an estimation of 
the timescale of tidal  friction for each component.  

\subsection {Calculation of gravitational potential energy $\Omega$ and 
          moment of inertia $I$} 

As indicated by Claret (2019) the effects of  General Relativity  on 
the calculation of the moment  of inertia and gravitational 
potential energy can be neglected  for  stars during the  PMS, 
 main sequence,  and even for WD. However,  for consistency with 
our previous papers on compact stars,  here we adopt the 
relativistic formalism throughout. Therefore, the moment of inertia can be 
computed using the      equations     

\begin{eqnarray}
{J = {{8\pi}\over{3}}\int_{0}^{R} \Lambda(r)r^4\left[\rho'(r) +
        P(r)/c^2\right] dr }, \nonumber\\
I \approx  {J\over{\left(1 + {2GJ\over{R^3c^2}}\right)}} \equiv
{(\beta R)^2}M, 
\end{eqnarray}

\noindent
where  $\beta$ is the radius of gyration.

The gravitational energy of a spherically symmetric star can be written as

\begin{eqnarray}
{\Omega = -4\pi\int_{0}^{R} {r^2\rho'(r)\left[\Lambda^{1/2}(r) - 1\right]dr} } 
\equiv -{\alpha} {G M^2\over{R}}.  
\end{eqnarray}

\noindent
In the above equation  $P(r)$ is the pressure,  $\rho'(r)$ the energy density, and 
the function $\Lambda(r)$ is given by  $\left[1 - {2 G m(r)\over{r c^2}}\right]^{-1}$. 
The parameter $\alpha$ is a dimensionless number that measures the relative mass 
concentration. In the case of less elaborated stellar models (e.g.  polytropes),  we have 
$\alpha = 3/(5-n)$, where $n$ is the polytropic index. Equations 9  and  10  were 
integrated simultaneously adopting the same numerical scheme and tolerance level 
as in   Eq. 1.

\subsection{Some interesting  properties of the moment of inertia and the gravitational           
potential energy: The $\Gamma$ function}

\begin{figure}
        \includegraphics[height=9cm,width=6cm,angle=-90]{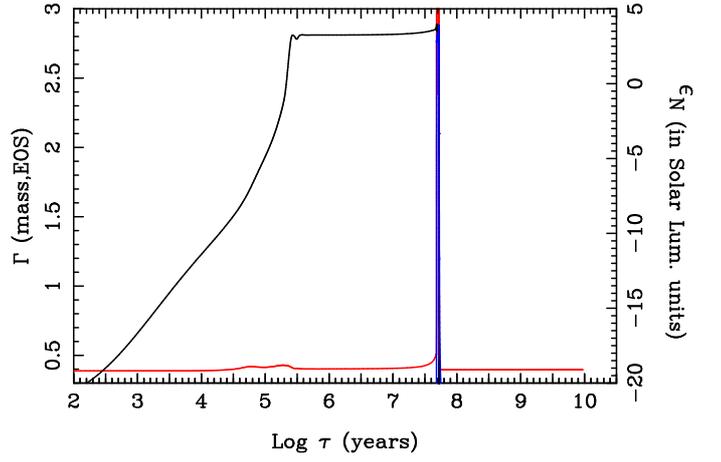}
        \caption{Time evolution of the function  $\Gamma$(mass, EOS) for a model 
                with initial mass of 7.00 M$_{\odot}$ evolving from the PMS to the WD stage,  
                $\alpha_{\rm MLT}$=1.84, f$_{ov}$ = 0.016,  [Fe/H]=0.00. The red line  
                represents  $\Gamma$(mass, EOS), while the black line indicates the 
                total thermal power from PP and CNO (excluding neutrinos)   and the blue line 
                indicates the total thermal power from triple-$\alpha$ (also excluding 
                neutrinos). The nuclear power $\epsilon_N$ is in   logarithmic scale.}
\end{figure}

\begin{figure}
        \includegraphics[height=9cm,width=6cm,angle=-90]{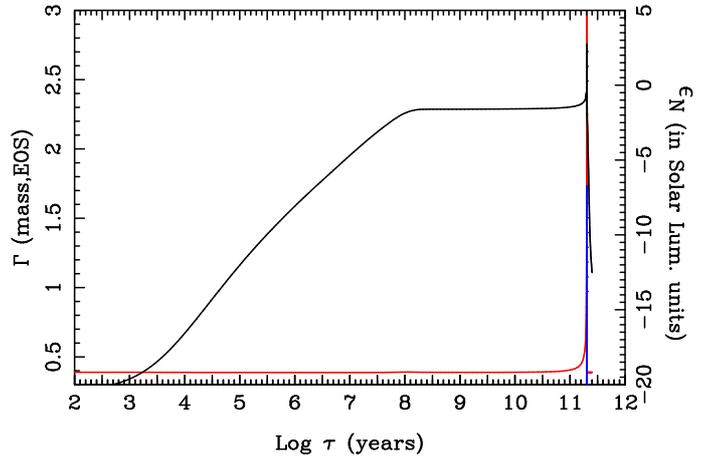}
        \caption{Same as   Figure 4, but  for a model with initial mass of 0.40 
        M$_{\odot}$ and  f$_{ov}$ = 0.000.}
\end{figure}

 The factors  $\alpha$ and $\beta$  are connected through the function  $\Gamma$ 
 introduced by Claret (2012) and improved by Claret\&Hempel (2013), which is defined  as 

\begin{eqnarray} 
\Gamma(mass, EOS) \equiv {\left[\alpha\beta\right]\over{\Lambda(R)^{0.8}}},   
\end{eqnarray}

\noindent
where EOS is the   equation of state. 

One of the most striking properties of this function is that the final products of 
stellar evolution (white dwarfs, neutron-quark hybrids,  and proto-neutron stars at the onset of formation of black holes)   
recover the value calculated for the PMS stage (i.e.  $\Gamma$(mass, EOS)) $\approx$ 0.40. 
We note   for the last four  mentioned systems that the effects of  General 
Relativity are strong. This invariance was also extended to models of gaseous planets 
with masses between 0.1 and 50.0 M$_{\rm Jupiter}$,  following from the gravitational contraction 
up to an age of $\approx$ 20 Myr. As a consequence of this 
invariance a macroscopic stability criterion for neutron, hybrid, and quark star 
models was established. More detailed information on this function, the  
`memory effect', and the stability criterion can be found in Claret (2012), 
Claret $\&$Hempel (2013), and Claret (2014).

As examples of the behaviour of  $\Gamma$(mass, EOS),  Figures 4 and 5 show the invariance of such a  
function for the PMS-WD stages for two different models, 7.00 and 0.40 M$_{\odot}$, respectively, 
adopting the solar composition. In both figures $\Gamma$(mass, EOS) increases by about three orders of 
magnitude with respect to its value at PMS (this increase is not shown fully in Figs. 4 and 5 due 
to the chosen scale). 

On the other hand, it is clear from both figures    that there is a connection between the values of 
$\Gamma$(mass, EOS) and the total thermal power from PP, CNO, and triple-$\alpha$: the larger 
the thermonuclear contributions, the larger the value of $\Gamma$(mass, EOS). During the PMS phase, 
when the chemical composition is homogeneous,  $\Gamma$(mass, EOS) $\approx$ 0.40 and $\epsilon_{\rm N}$ 
$\approx$ 0.0. 
However, in the WD phase,  although the initial chemical composition has been altered by 
  thermonuclear reactions,  the value 0.40 is recovered, given that these reactions cease.  
In summary,  the property of $\Gamma$(mass, EOS)    
 presents the same value ($\approx$ 0.40) in the initial and final stages of stellar
 evolution. We     confirmed this behaviour for all  the models of 
 our grids.  This behaviour is  also valid for gaseous planets, neutron-quark-hybrid   stars, and 
 proto-neutron stars at the onset of formation of black holes.

\section{Final remarks and table organization} 

We   computed three evolutionary grids covering three metallicities: [Fe/H)=-0.50, 0.00, 
and +0.50 from PMS to the WD stage. The covered mass range was 0.20--8.00  M$_{\odot}$. For such 
models, in addition to the characteristic parameters (age, luminosity, log g, effective 
temperatures), the internal structure constants (k$_2$, k$_3$, k$_4$), the moment of inertia, 
and the gravitational potential energy have also been computed. 
The resulting tables have been prepared  mainly for  studies of DLEBS and/or exoplanetary systems.  
Tables I-III summarize the input physics for each series of models, while  Tables 1--69 contain 
the necessary theoretical inputs for the comparison with the absolute dimensions of the DLEBS  
as well as the necessary parameters for the apsidal motion and tidal evolution studies.

\begin{acknowledgements} 
 I thank M. Broz for his pertinent comments and suggestions that have improved this paper.
The Spanish MEC (AYA2015-71718-R and ESP2017-87676-C5-2-R) is gratefully acknowledged 
for its support during the development of this work. AC also acknowledges financial 
support from the grant CEX2021-001131-S funded by MCIN/AEI/10.13039/501100011033. 
This research has made use of the SIMBAD database, operated at the CDS,  Strasbourg, 
France, and of NASA's Astrophysics Data System Abstract Service.
\end{acknowledgements}

{}

\begin{appendix}

\section{Brief description of Tables A1, A2, and A3}
Tables A1, A2, and A3 summarize the type 
of data available (for more details, see the ReadMe file 
on the CDS).

\renewcommand{\tablename}{Table }
\begin{table*}
        \caption{Mass and initial chemical composition}
        \begin{flushleft}
                \centering
                \begin{tabular}{ccc}                         
                        \hline                         
                        Name    & Initial mass  (M$_{\odot}$)   &  [Fe/H]  \\ 
                        \hline                  
                        Table1    &0.20  & 0.00\\
                        Table2    &0.30  & 0.00\\
                        Table3    &0.40  & 0.00\\
                        Table4    &0.50  & 0.00\\
                        Table5    &0.60  & 0.00\\
                        Table6    &0.70  & 0.00\\
                        Table7    &0.80  & 0.00\\
                        Table8    &0.90  & 0.00\\
                        Table9    &1.00  & 0.00\\
                        Table10   &1.10  & 0.00\\
                        Table11   &1.20  & 0.00\\
                        Table12   &1.40  & 0.00\\
                        Table13   &1.60  & 0.00\\
                        Table14   &1.80  & 0.00\\
                        Table15   &2.00  & 0.00\\
                        Table16   &2.50  & 0.00\\
                        Table17   &3.00  & 0.00\\               
                        Table18   &4.00  & 0.00\\
                        Table19   &5.00  & 0.00\\
                        Table20   &6.00  & 0.00\\
                        Table21   &7.00  & 0.00\\
                        Table22   &7.50  & 0.00\\
                        Table23   &8.00  & 0.00\\
                        \hline
                \end{tabular}
        \end{flushleft}
\end{table*}

\renewcommand{\tablename}{Table }
\begin{table*}
        \caption{Mass and initial chemical composition}
        \begin{flushleft}
                \centering
                \begin{tabular}{ccc}                         
                        \hline                         
                        Name    & Initial mass  (M$_{\odot}$)   &  [Fe/H]  \\ 
                        \hline  
Table24   &0.20  &-0.50\\
Table25   &0.30  &-0.50\\
Table26   &0.40  &-0.50\\
Table27   &0.50  &-0.50\\
Table28   &0.60  &-0.50\\
Table29   &0.70  &-0.50\\
Table30   &0.80  &-0.50\\
Table31   &0.90  &-0.50\\
Table32   &1.00  &-0.50\\
Table33   &1.10  &-0.50\\
Table34   &1.20  &-0.50\\
Table35   &1.40  &-0.50\\
Table36   &1.60  &-0.50\\
Table37   &1.80  &-0.50\\
Table38   &2.00  &-0.50\\
Table39   &2.50  &-0.50\\
Table40   &3.00  &-0.50\\               
Table41   &4.00  &-0.50\\
Table42   &5.00  &-0.50\\
Table43   &6.00  &-0.50\\
Table44   &7.00  &-0.50\\
Table45   &7.50  &-0.50\\
Table46   &8.00  &-0.50\\               

\hline
\end{tabular}
\end{flushleft}
\end{table*}

\renewcommand{\tablename}{Table }
\begin{table*}
        \caption{Mass and initial chemical composition}
        \begin{flushleft}
                \centering
                \begin{tabular}{ccc}                         
                        \hline                         
                        Name    & Initial mass  (M$_{\odot}$)   &  [Fe/H]  \\ 
                        \hline  
            Table47   &0.20  & +0.50\\
Table48   &0.30  & +0.50\\
Table49   &0.40  & +0.50\\
Table50   &0.50  & +0.50\\
Table51   &0.60  & +0.50\\
Table52   &0.70  & +0.50\\
Table53   &0.80  & +0.50\\
Table54   &0.90  & +0.50\\
Table55   &1.00  & +0.50\\
Table56   &1.10  & +0.50\\
Table57   &1.20  & +0.50\\
Table58   &1.40  & +0.50\\
Table59   &1.60  & +0.50\\
Table60   &1.80  & +0.50\\
Table61   &2.00  & +0.50\\
Table62   &2.50  & +0.50\\
Table63   &3.00  & +0.50\\              
Table64   &4.00  & +0.50\\
Table65   &5.00  & +0.50\\
Table66   &6.00  & +0.50\\
Table67   &7.00  & +0.50\\
Table68   &7.50  & +0.50\\
Table69   &7.50  & +0.50\\      
                        
                        \hline
                \end{tabular}
        \end{flushleft}
\end{table*}

\end{appendix}
\end{document}